\documentclass[%
 reprint,
 amsmath,amssymb,
 aps,
]{revtex4-1}
\usepackage{silence}
\usepackage[utf8]{inputenc}
\usepackage{graphicx}
\usepackage{dcolumn}
\usepackage{bm}
\usepackage{ulem}
\usepackage{xcolor}
\usepackage{lipsum}
\usepackage{hyperref}
\hypersetup{
    colorlinks=true,
    linkcolor=blue,
    filecolor=magenta,      
    urlcolor=black,
    citecolor=blue,
    pdftitle={Overleaf Example},
    pdfpagemode=FullScreen,
    }

\begin{document}

\preprint{APS/123-QED}

\title{Self-referenced nonlinear interferometry for chromatic dispersion sensing across multiple length scales}

\author{Romain Dalidet, Sébastien Tanzilli, Gregory Sauder, Laurent Labonté}
\email{laurent.labonte@univ-cotedazur.fr}
\author{Anthony Martin}
\affiliation{Université Côte d’Azur, CNRS, Institut de physique de Nice, France}

\begin{abstract}
Chromatic dispersion critically impacts the performance of numerous applications ranging from telecommunication links to ultrafast optics and nonlinear devices, yet fast and precise measurements are challenging, especially for short length–dispersion products. We present a fully fiber-integrated nonlinear Sagnac interferometer that exploits cascaded second-order processes to generate frequency- anticorrelated idler light and achieve odd-order dispersion cancellation without active stabilization. The measurement is intrinsically self-referenced, as the dispersion-induced phase is extracted from the interference between counter-propagating nonlinear processes within the same Sagnac loop, eliminating the need for an external reference arm or prior calibration.
Operating entirely at telecom wavelengths and read out on a standard optical spectrum analyzer, the device produces instantaneous, high-visibility fringes and calibration-free spectra using dual-port normalization. We demonstrate chromatic dispersion measurements on fiber samples ranging from 25 cm to 4 km, spanning short fiber segments to long-haul links. This architecture combines self-stability, broadband compatibility, and rapid acquisition, offering a practical metrology tool for both research and industry.
 \end{abstract}

\maketitle

\section{Introduction}

Optical phase estimation is a cornerstone of both classical and quantum metrology, enabling high-performance measurements in fields as diverse as gravitational-wave detection \cite{Abbott2016}, optical coherence tomography \cite{Huang1991OCT}, displacement sensing \cite{Xu2013displacement}, and gyroscopy \cite{Fink_2019}. In applied photonics, it underpins the characterization of waveguide dispersion \cite{Dalidet2023, Baker2014} and the development of integrated photonic circuits \cite{Bogaerts2012}.

Accurate chromatic dispersion (CD) measurements are particularly challenging when the length–dispersion product of the sample is small, as in the case of short fibers, integrated photonic devices, or specialty components. Conventional phase-shift or modulation-based techniques \cite{Baker2014, Abedin2000} lose sensitivity in this regime, whereas classical white-light interferometry requires precise wavelength balancing \cite{Kaiser2018}. In industrial metrology, there is a growing demand for methods that combine broadband coverage, high precision, and short acquisition times without the complexity of environmental isolation or free-space alignment.

Classical interferometers such as Mach–Zehnder and Michelson designs are widely used for precision phase sensing, but they are highly sensitive to environmental fluctuations and typically require active stabilization \cite{Kaiser2018}. Nonlinear interferometers, including SU(1,1) configurations \cite{Hudelist2014, chekhova2016} and quantum-induced-coherence architectures \cite{Kalashnikov2016,Lemos2014,Paterova_2018}, have been applied to CD measurement \cite{Riazi2019,Riazi2020}, offering intrinsic odd-order dispersion cancellation \cite{Franson1992} and relaxed path-length stability requirements. However, many implementations require the sample under test to be transparent at two wavelengths, which often limits the experiment  to bulk optics and alignment-sensitive setups. More generally, quantum-probe operations, such as two-photon interference, rely on narrowband filtering and single-photon detection, resulting in long acquisition times.

Here, we introduce a fully fibered, self-referenced nonlinear interferometer based on cascaded second-order processes in a Sagnac loop to accurately measure the CD. This design achieves self-stabilization \cite{Kim2006} and odd-order dispersion cancellation while operating entirely at telecom wavelengths. This architecture allows direct broadband readout on a standard optical spectrum analyzer, enabling fast, calibration-free measurements from short to long fiber samples, effectively filling a gap in interferometric sensing and meeting the needs of industrial metrology.

\section{Methods}

\begin{figure*}[!ht]
    \centering
    \includegraphics[width=0.9\linewidth]{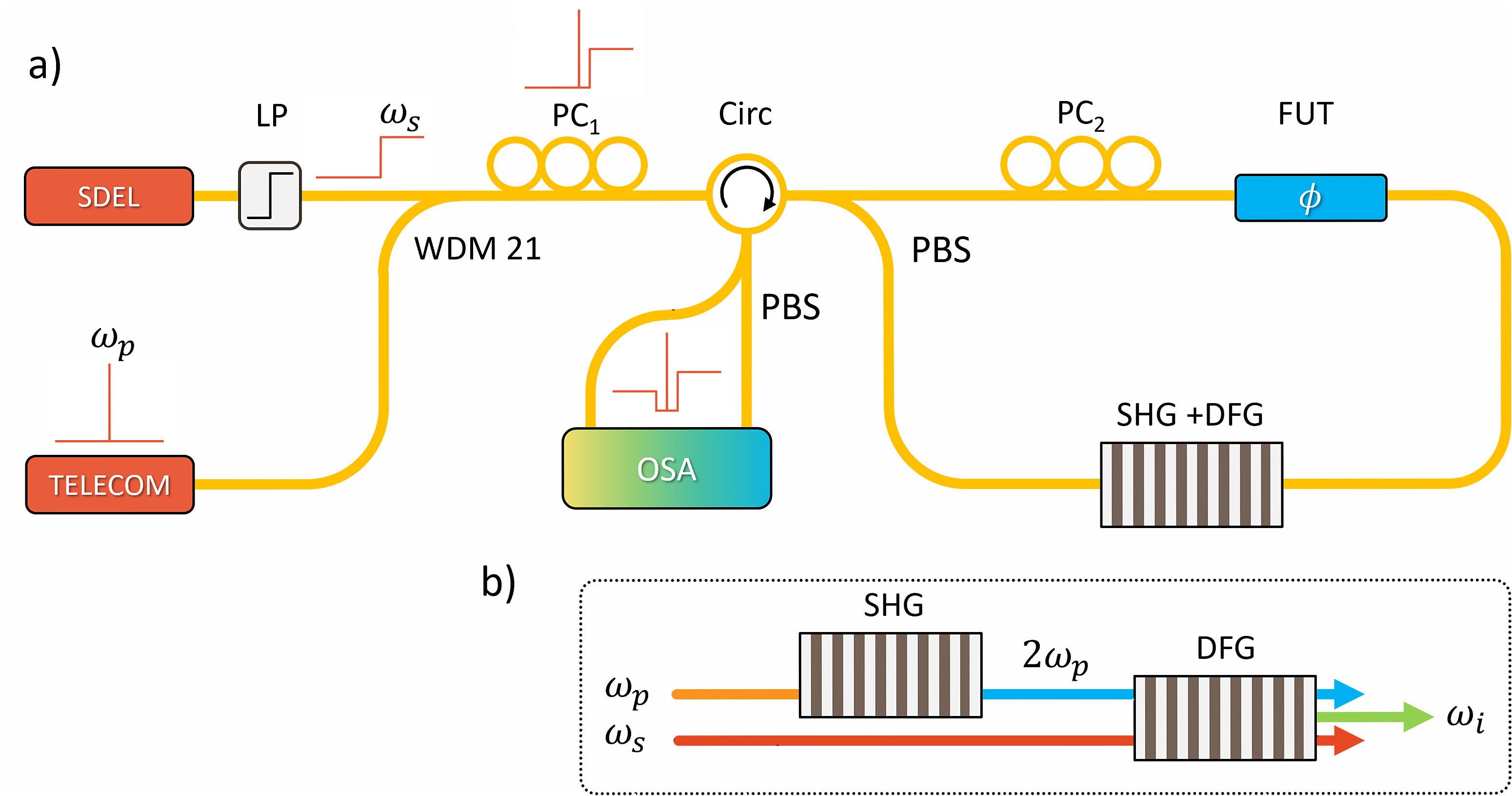}
    \caption{a) Experimental setup for CD measurement using a cascaded nonlinear Sagnac interferometer. A Superluminescent diode and pump laser generate a frequency-anticorrelated superposed state inside the loop via a nonlinear crystal. The CD of the FUT is retrieved by measuring the interference fringes on a standard OSA. b) Schematic of the cascaded nonlinear process. The pump is first converted into its second harmonic, which then interacts with the signal in the second step to generate an idler via difference-frequency generation. SDEL : superluminescent diode. WDM: Wavelength-division multiplexer. PC: Polarization controller. Circ: circulator. PBS: Polarizing beam splitter. FUT : fiber under test. SHG: Second-harmonic generation. DFG: difference-frequency generation. OSA: optical spectrum analyzer.}
    \label{fig: experiment}
\end{figure*}

The dispersion of an optical material, or equivalently, the relative spectral phase in an interferometer, is commonly expressed as a Taylor expansion of the wave vector around the central frequency $\omega_0$ of the light propagating through it:
\begin{equation}\label{eq: taylor disp}
    k(\omega) =  \sum_{n=0}^{\infty} \left.\frac{\partial^n k}{\partial \omega^n}\right|_{\omega_0} \frac{(\omega - \omega_0)^n}{n!}   =  \sum_{n=0}^{\infty} \beta^{(n)} \frac{\Delta\omega^n}{n!} \, .
\end{equation}
Here, the second-order term $\beta^{(2)}$ corresponds to the chromatic dispersion. Because the hierarchy $\beta^{(n)} \gg \beta^{(n+1)}$ generally holds for optical materials, interferometric extraction of the CD in the frequency (or time) domain requires the cancellation of the first-order term $\beta^{(1)}$ (group delay) in the interferometer’s relative phase. 
In white-light interferometry, this is achieved by balancing the interferometer at the stationary-phase point so that the relative spectral phase becomes independent of $\beta^{(1)}$. In quantum white-light (QWLI) interferometry based on energy-correlated photon pairs, energy and phase conservation impose that the total phase is the sum of the individual phases; under these constraints, all odd-order dispersion terms cancel in the relative phase (nonlocal dispersion cancellation) \cite{Franson1992}. Importantly, although this cancellation was originally interpreted as a nonlocal quantum effect, the suppression of odd-order dispersion terms is, from a local perspective, a direct consequence of energy and phase conservation during the nonlinear generation process \cite{Dalidet2025}. This local interpretation forms the basis of our experimental approach, as illustrated in Fig. ~\ref{fig: experiment}(a), respectively.

\medskip
A telecom laser centered at $\lambda_p = 1560.6$~nm and a superluminescent diode (SDEL) filtered in the L-band using a long-pass filter (LP) are used as pump sources. The frequency of the filtered SDEL, denoted $\omega_s$, defines the signal beam. Both the pump and signal are combined via a wavelength division multiplexer (WDM) and transmitted through a polarization controller \( \mathrm{PC}_1 \) and a circulator. The first output port of the circulator feeds a nonlinear Sagnac interferometer composed of a polarizing beam splitter (PBS), second PC, fiber under test (FUT), and polarization-maintaining nonlinear crystal (MgO-doped LiNbO\textsubscript{3}) engineered for second-order nonlinear interactions. We emphasize that both outputs of the PBS are aligned along the same axis and that all fibers, including those of the crystal, are PM fibres. Therefore, if the FUT is also a PM, the entire Sagnac loop preserves the polarization, and the PC inside the loop can be removed.

First, we describe the cascaded nonlinear interaction in the crystal, as shown schematically in Fig. ~\ref{fig: experiment}(b). For simplicity, we considered a process occurring in two identical nonlinear crystals. The first crystal, pumped by the telecom laser, performs second-harmonic generation (SHG) at a frequency $2\omega_p$. The second crystal, pumped by this SHG beam and signal $\omega_s$, performs difference-frequency generation (DFG), producing an idler at frequency $\omega_i$ satisfying

\begin{align}
    \omega_i &= 2\omega_p - \omega_s \, , \label{eq: energy conservation idler} \\ 
    k_i &= 2k_p - k_s \, . \label{eq: phase conservation idler}
\end{align}
A factor of 2 in these expressions arises from the SHG process, whereas the minus sign corresponds to the DFG step. Thus, the idler is energy-anticorrelated with the signal, centered around the pump frequency $\omega_p$. In our experiment, both SHG and DFG occur in the same crystal designed for operation in the C+L band, which is phase-matched at $2\omega_p$, and the poling period and phase-matching temperature are identical for both nonlinear steps. Because the SHG beam is only an intermediate, it is not collected; the pigtails of the crystal are optimized for telecom wavelengths to maximize the coupling and collection of the pump, signal, and idler beams.

\begin{figure}[!ht]
    \centering
    \includegraphics[width=\linewidth]{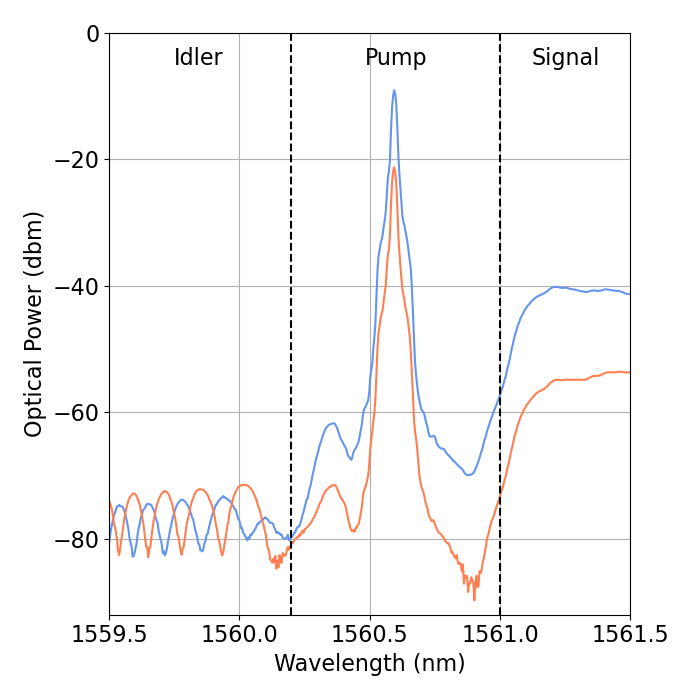}
    \caption{Raw spectra measured by the optical spectrum analyzer at the two output ports of the final PBS in the setup. The complementary spectral modulations in the idler part arise from the interference and confirm the high visibility and intrinsic stability of the Sagnac configuration.}
    \label{fig: raw data}
\end{figure}
We now turn our attention to the interferometric configuration. The Sagnac loop provides intrinsic phase stability, for any external perturbations, for example thermal or mechanical, evolving on time scales much slower than the loop round-trip, the relative phase between clockwise (CW) and counterclockwise (CCW) paths remains zero. In the absence of rotation, both paths are equal in length, relaxing the usual coherence time constraints on the pump sources. This self-referencing property explains the widespread use of Sagnac interferometers to probe non-reciprocal phase shifts, such as those induced by the Faraday or Sagnac effects. To measure chromatic dispersion while preserving phase stability, the interferometer must contain a non-reciprocal element that encodes the sample dispersion differently in the two directions. In this case, the nonlinear crystal fulfils this role. By preparing both the pump and signal beams in a diagonal polarization state at the loop entrance, we generate a balanced superposition of CW and CCW paths. In the CW arm, the pump and signal propagate through the FUT before entering the crystal, whereas in the CCW arm, the nonlinear interaction occurs first, and the idler then propagates through the FUT. This asymmetry produces a non-reciprocal spectral phase difference.

As previously stated, in the CW arm, the pump and signal pass through the FUT before entering the crystal and generating an idler. The resulting phase is:
\begin{equation}\label{eq: phase CW}
    \phi_{\mathrm{CW}} = (2k_p - k_s)L \, ,
\end{equation}
where $L$ denotes the length of the FUT. In the CCW arm, a nonlinear interaction occurs first, generating the idler, which then propagates through the FUT:
\begin{equation}\label{eq: phase CCW}
    \phi_{\mathrm{CCW}} = k_i L \, .
\end{equation}

The resulting non-reciprocal relative phase is therefore
\begin{equation}\label{eq: relative phase}
    \begin{split}
        \Phi &= \phi_{\mathrm{CW}} - \phi_{\mathrm{CCW}} \\
         &=  (2k_p - k_s - k_i)L \, .
    \end{split}
\end{equation}
Using energy conservation for the cascaded SHG–DFG process, the frequency detuning of the signal and idler are anticorrelated $\Delta\omega_s = -\Delta\omega_i = \Delta\omega$. Substituting Eq. \ref{eq: taylor disp} into Eq. \ref{eq: relative phase} yields:
\begin{equation}\label{eq: non local disp cancel}
 \Phi =   \beta^{(2)} L \, \Delta\omega^2 +  \mathcal{O}(\Delta\omega^4) \, .
\end{equation}

This expression shows that all odd-order dispersion terms vanish, thereby demonstrating nonlocal dispersion cancellation. Furthermore, because all beams propagate through the same Sagnac loop, the zeroth-order phase term also cancels naturally. As a result, the interference pattern depends solely on the chromatic dispersion of the fiber under test, which enhances the measurement precision by reducing the number of relevant parameters. This property also underpins the self-referenced nature of the measurement: the relative phase is not defined with respect to an independent reference arm, but arises directly from the interference between the two counter-propagating nonlinear contributions generated within the same loop. Consequently, the extracted dispersion is inherently insensitive to absolute path lengths, global phase offsets, and common-mode environmental fluctuations.
Finally, all the light exits the loop via the PBS and is directed to the final port of the circulator. A second PBS then projects the idler interference onto the appropriate polarization basis, and both output ports are recorded on a commercial optical spectrum analyzer (OSA), as shown in Fig.~\ref{fig: raw data}. This fully fiber-based readout provides fast, alignment-free chromatic dispersion measurements without the need for active synchronization or narrowband filtering, in contrast to conventional QWLI schemes.

\section{Discussion}

The probability of detecting an idler photon at the output of the interferometer is given by
\begin{equation}\label{eq: detection proba}
    P_{H,V}(\omega_i) \propto \frac{S(\omega_p,\omega_s,\omega_i)}{2} \left( 1 \pm \mathcal{V} \cos\Phi \right) \, ,
\end{equation}
where \( S \) is the total transmission spectrum of the system, encompassing the DFG spectral acceptance of the nonlinear crystal and the transmission of all optical components. The subscripts $H$ and $V$ denote the horizontal and vertical polarization, respectively, and $\mathcal{V}$ is the fringe visibility.

Because both outputs of the interferometer are measured, the normalized interference pattern reads:
\begin{equation}\label{eq: normalized proba}
    P = \frac{P_H - P_V}{P_H + P_V} = \mathcal{V} \cos\Phi \, .
\end{equation}
This normalization eliminates the need for prior calibration of the system transmission and suppresses the influence of losses and spectral response. The fringe contrast quantifies the indistinguishability of the two idler contributions in the loop and is governed by the following conditions:

\begin{itemize}
    \item Polarization alignment along the loop: If the FUT is polarization-maintaining (PM), the loop preserves polarization without further adjustment. Otherwise, a polarization controller (\( \mathrm{PC}_2 \), see Fig.~\ref{fig: experiment}.a) is used to match the states in the two arms; in this case, an initial calibration of the CD with \( \mathrm{PC}_2 \) is required.
    
    \item Spectral overlap of the CW and CCW contributions: This condition is inherently satisfied since only one nonlinear crystal is used, ensuring identical nonlinear optical properties in both directions.
    
    \item Balance of idler spectral amplitudes : Losses in the FUT or asymmetric coupling at the crystal pigtails can introduce imbalance. This can be compensated for by adjusting the input pump and signal power ratio with \( \mathrm{PC}_1 \) (see Fig.~\ref{fig: experiment}.a).
\end{itemize}
\onecolumngrid

\begin{figure}[!h]
    \centering
    \includegraphics[width=1\linewidth]{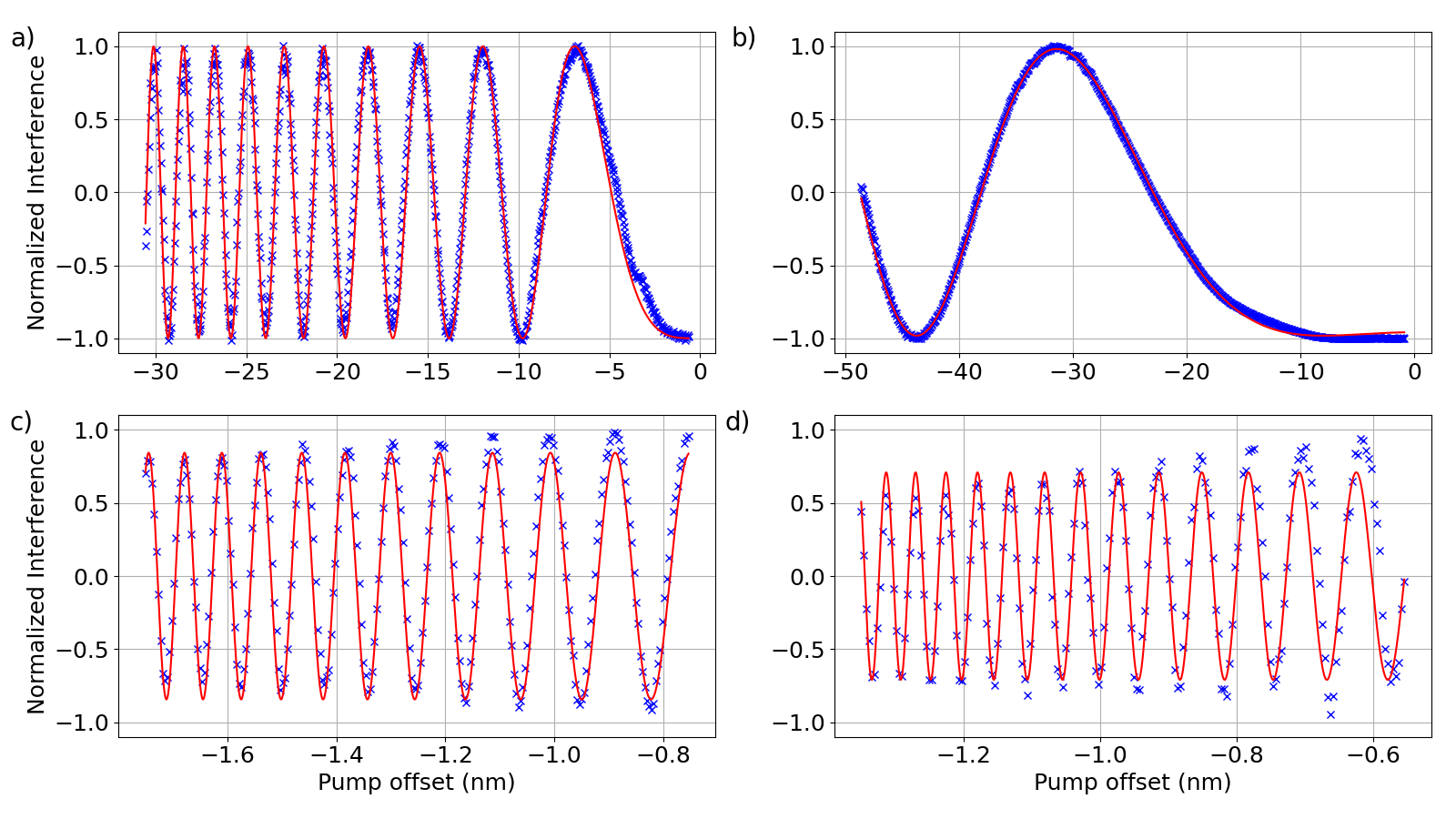}
    \caption{Normalized interference patterns (experimental data, blue) and corresponding fits (red) using Eq.~\eqref{eq: normalized proba}. The results are shown for (a)  a 1m long PM dispersion-compensating fiber, (b) a 25cm long PM fiber, (c) a dispersion-compensation fiber module of \( L \approx 200~\mathrm{m} \), and (d) a \( L \approx 4~\mathrm{km} \) coil of standard telecom single-mode fiber. These measurements illustrate the versatility of the method, covering dispersion–length products from sub-meter to multi-kilometer scale.}
    \label{fig: full results}
\end{figure}
\twocolumngrid
Two additional considerations are critical for optimizing CD measurements. First, the residual background noise originating from the pump and signal beams in the idler detection band must be properly evaluated. This noise differs depending on the output port owing to the intrinsic phase stability of the loop. Second, care must be taken to prevent the SLED from producing second-harmonic light. If so, the energy and phase conservation conditions for the idler would no longer reference a fixed frequency, thereby invalidating the odd-order dispersion cancellation. In our case, the full width at half maximum (FWHM) of the crystal’s SHG acceptance bandwidth is narrower than the WDM bandwidth used to combine the pump and signal, effectively preventing unwanted SHG. 

The first CD measurement, shown in Fig. \ref{fig: full results} (a) was performed with a 1~m long PM dispersion-compensating fiber. Fitting the normalized interference pattern with Eq. ~\eqref{eq: normalized proba} yields a CD value of \( 8.200 \times 10^{-2} \pm 3 \times 10^{-5}~\mathrm{ps/nm} \), which is among the most precise interferometric CD measurements reported to date \cite{Kaiser2018, Riazi2020, Dalidet2023}. The high fringe visibility (\( \mathcal{V} \approx 95\% \)) confirms the validity of the protocol. The residual CD measurement, i.e. with no fiber under test, does not exhibit interference fringes over the entire spectral range.\\
To demonstrate the versatility of the method, we extended the measurement to three additional samples: (i) a 25cm long PM fiber, (ii) a fiber-based dispersion compensation module of length \( L \approx 200~\mathrm{m} \) and (iii) a fiber coil of \( \approx 4~\mathrm{km} \). The corresponding normalized spectra and fits are shown in Fig. ~\ref{fig: full results}, yielding CD values of \( 4.128 \times 10^{-3} \pm 3 \times 10^{-6}~\mathrm{ps/nm} \), \( 35.97 \pm 1 \times 10^{-2}~\mathrm{ps/nm} \) and \( 72.95 \pm 4 \times 10^{-2}~\mathrm{ps/nm} \), respectively. For longer samples (Fig. \ref{fig: full results} (c) and (d)), the fringe visibility decreases slightly as the measurement deviates from the pump frequency. This effect arises from the narrowing fringe period approaching the resolution limit of the OSA, \( \Delta\lambda = 4~\mathrm{pm} \). From this limitation, we estimate that the setup can resolve fringes for CD values up to \( 170~\mathrm{ps/nm} \), equivalent to \( \approx 10~\mathrm{km} \) of standard single-mode telecom fiber. Conversely, for short-length dispersion products, the minimum measurable CD depends on the criterion adopted for resolving interference fringes. If we consider the conservative requirement of observing at least one full fringe, its spectral width is
\begin{equation}\label{eq: fringe bandwidth}
    \Delta\lambda = \left( \frac{\lambda_p^4}{2\pi c^2 |\beta^{(2)}| L} \right)^{1/2} \, .
\end{equation}
Using \( \Delta\lambda \approx 90~\mathrm{nm} \), corresponding to the FWHM of the crystal’s DFG acceptance, this criterion yields a minimum resolvable CD of \( \approx 10^{-3}~\mathrm{ps/nm} \)
, which is equivalent to approximately 6 cm of standard single-mode telecom fibre. Such sensitivity makes the setup particularly relevant for characterizing very short devices, including integrated waveguides and nonlinear components, where dispersion plays a critical role. This minimum value can be further reduced by employing a shorter nonlinear crystal, which increases the DFG acceptance bandwidth.\\

Although the ultimate sensitivity of our experiment is bounded by the standard quantum limit, the proposed scheme removes many of the practical limitations encountered in previous interferometric approaches. Linear interferometers require passive or active stabilisation of the relative phase, which increases complexity and can restrict the accessible sample length. SU(1,1) nonlinear interferometers offer intrinsic phase sensitivity but impose strict constraints: the sample under test must be transparent to both pump and signal wavelengths, and the losses directly degrade the fringe visibility. In contrast, in our configuration, the pump analogue at frequency $2\omega_p$ is confined within the nonlinear crystal, so that only telecom-band light circulates in the loop, and the visibility can be restored by tuning the input state polarization. Quantum two-photon approaches further require spectral or time-domain filtering and single-photon detection, leading to long acquisition times and higher system complexity, whereas our method relies solely on bright classical fields and standard spectral detection. In addition, our architecture operates without coherence-time restrictions on the pump laser relative to the sample length, unlike many conventional schemes. Altogether, the combination of self-stabilization and self-referencing, full telecom compatibility, robustness to loss, and simple broadband readout makes our approach a practical and versatile alternative to existing interferometric techniques for chromatic dispersion measurements.

\section{Conclusion}

We have presented a new interferometric method for chromatic dispersion measurement based on a cascaded nonlinear Sagnac loop. By exploiting energy and phase conservation in the nonlinear process, the interference spectrum becomes dependent only on the chromatic dispersion of the sample, with the zeroth-order and all odd-order dispersion terms intrinsically cancelled. This allowed us to obtain high-contrast, calibration-free interference fringes using standard telecom components, achieving state-of-the-art precision. We further demonstrated the versatility of the method by characterizing samples ranging from tens of centimeters to several kilometers, showing that both small and large dispersion–length products can be accessed within the same platform. The combination of self-referencing, full telecom compatibility, broadband acceptance, and rapid acquisition makes this technique a practical and scalable solution for chromatic dispersion metrology in both laboratory and industrial environments.

\section*{Acknowledgment}

This work was conducted within the framework of the OPTIMAL project, granted by the European Union through the Fond Européen de développement régional (FEDER). The authors also acknowledge financial support from the Agence Nationale de la Recherche (ANR) through the projects METROPOLIS (ANR-19-CE47-0008), QAFEINE (21-ASTR-0007-DA), and PARADIS (ANR-22-ASTR-0027-01).\\

\section*{Author information}

All authors contributed equally to the entire process, from the first draft to the final version of the manuscript before submission. All authors have read, discussed, and contributed to the writing, reviewing, and editing of the manuscript. 

\section*{Competing interests}
The authors declare no conflict of interest.

\section*{Data Availability}
The data are available from the authors upon reasonable request.

\end{document}